\documentclass[useAMS,usenatbib]{mn2e}
\usepackage{natbib}

\usepackage[dvips]{graphics}
\usepackage{epsfig}
\usepackage{amsmath}
\usepackage{aas_macros}
\usepackage{hyperref}
\hypersetup{backref=true, pagebackref=true, hyperindex=true, breaklinks=true,colorlinks=true,urlcolor=blue, linkcolor=black,  citecolor=blue,pagecolor=red, bookmarks=true, bookmarksopen=true}

\def\mearth{{\rm\,M_\oplus}}

\def\msun{{\rm\,M_\odot}}
\def\deg{^\circ}

\title[]{Implications of the interstellar object 1I/'Oumuamua for planetary dynamics and planetesimal formation}
\author[Raymond et al]{Sean N. Raymond$^1$\thanks{E-mail: rayray.sean@gmail.com}, Philip J. Armitage$^{2,3}$, Dimitri Veras$^\mathrm{4,5}$\thanks{STFC Ernest Rutherford Fellow}, \newauthor Elisa V. Quintana$^6$, \& Thomas Barclay$^{6,7}$\\
$^\mathrm{1}$Laboratoire d'Astrophysique de Bordeaux, CNRS and Universit{\'e} de Bordeaux, All{\'e}e Geoffroy St. Hilaire, 33165 Pessac, France \\
$^2$JILA, University of Colorado and NIST, 440 UCB, Boulder, CO 80309-0440, USA\\
$^3$Department of Astrophysical \& Planetary Sciences, University of Colorado, Boulder, CO 80309-0391, USA\\
$^4$Department of Physics, University of Warwick, Coventry CV4 7AL, UK\\
$^5$Centre for Exoplanets and Habitability, University of Warwick, Coventry CV4 7AL, UK\\
$^6$NASA Goddard Space Flight Center, 8800 Greenbelt Rd, Greenbelt, MD 20771, USA\\
$^7$University of Maryland, Baltimore County, 1000 Hilltop Cir, Baltimore, MD 21250, USA
}

\begin{document}

%\date{Nov 9, 2017}

\pagerange{\pageref{firstpage}--\pageref{lastpage}} \pubyear{2017}

\maketitle

\label{firstpage}

\begin{abstract}
'Oumuamua, the first bona-fide interstellar planetesimal, was discovered passing through our Solar System on a hyperbolic orbit. This object was likely dynamically ejected from an extrasolar planetary system after a series of close encounters with gas giant planets. To account for 'Oumuamua's detection, simple arguments suggest that $\sim 1 \ M_\oplus$ of planetesimals are ejected per Solar mass of Galactic stars.  However, that value assumes mono-sized planetesimals. If the planetesimal mass distribution is instead top-heavy the inferred mass in interstellar planetesimals increases to an implausibly high value. The tension between theoretical expectations for the planetesimal mass function and the observation of 'Oumuamua can be relieved if a small fraction ($\sim 0.1-1\%$) of planetesimals are tidally disrupted on the pathway to ejection into 'Oumuamua-sized fragments.  Using a large suite of simulations of giant planet dynamics including planetesimals, we confirm that 0.1-1\% of planetesimals pass within the tidal disruption radius of a gas giant on their pathway to ejection. 'Oumuamua may thus represent a surviving fragment of a disrupted planetesimal. Finally, we argue that an asteroidal composition is dynamically disfavoured for 'Oumuamua, as asteroidal planetesimals are both less abundant and ejected at a lower efficiency than cometary planetesimals.  
%Given the uncertainty in various parameters, the ejected mass inferred for mono-sized planetesimals may underestimate or overestimate the true mass.
\end{abstract}


\section{Introduction}
1I/'Oumuamua was discovered by the Panoramic Survey Telescope And Rapid Response System (Pan-STARRS) survey~\citep{chambers16} on Oct 19 2017~\citep{meech17,williams17}.  The object's orbit was determined to be inclined by 123$\deg$ with respect to the ecliptic plane and to have an eccentricity of 1.2, placing it on a hyperbolic orbit that is not aligned with any known cluster of comets~\citep{delafuente17}.  Given the lack of Solar System objects capable of imparting a large enough impulse to explain its anomalous velocity~\citep[e.g.][]{portegieszwart17}, 'Oumuamua is likely of extra-solar origin and in the process of passing through the Solar System on its way back to interstellar space.   

Interstellar planetesimals are likely to represent by-products of planet formation. Their existence is not a surprise as planet formation is well known to be less than 100\% efficient. The detectability of interstellar asteroidal and cometary planetesimals has been studied by a number of authors over the past several decades~\citep[e.g.][]{mcglynn89,kresak92,jewitt03,moromartin09,jura11,cook16}.  The absence of detections prior to 'Oumuamua led \cite{engelhardt17} to place an upper limit of $1.4 \times 10^{-3}$ AU$^{-3}$ on the abundance of interstellar planetesimals.

Planetesimals are thought to form via gas-assisted concentration of drifting particles in planet-forming discs~\citep[see the review by][]{johansen14}.  A fraction of planetesimals can subsequently be ejected by receiving gravitational kicks from planets. A planet's ejection potential can be quantified with the Safronov number $\Theta$, defined as the ratio of the escape speed from a planet's surface to the local escape speed from the star~\citep[see, e.g.,][]{ford08,raymond10,wyatt17}. Ejections are likely for $\Theta \gg 1$, while collisions are favored over ejections if $\Theta < 1$. In the Solar System, although all the giant planets have $\Theta > 1$, Jupiter ejects the vast majority of planetesimals. Given the planets' orbital architecture, outer Solar System planetesimals are generally passed inward from the ice giants to Saturn to Jupiter before being ejected after repeated encounters with Jupiter~\citep{fernandez84}. The resulting orbital migration is the basis for understanding Pluto's orbit \citep{malhotra95}, the Kuiper Belt~\citep[e.g.][]{nesvorny15}, and the current architecture of the outer Solar System \citep{tsiganis05}. Contemporary models for early outer Solar System evolution invoke planetesimal discs with masses of $\approx 15-65 \ M_\oplus$ \citep{deienno17}, most of which is ultimately ejected. 

Dynamical arguments show that the observation of one interstellar object during the duration of the Pan-STARRS survey implies that every star ejects about $1 \ M_\oplus$ of planetesimals~\citep{laughlin17,trilling17}. By comparing this to the $>10 \mearth$ ejected from the Solar System, one might immediately conclude that the observation of 'Oumuamua is expected and possibly overdue. However, this expectation is not obvious. Most of the known exoplanets are too close to their stars to be good candidates for ejecting planetesimals~\citep{laughlin17}, and the known population of gas giants that probably dominate ejections are present around only a small fraction of stars. Planetesimals, moreover, have a distribution of masses, and if that distribution is skewed toward large masses inferences from seeing a single small body need to be corrected for the much larger mass of the population. 

Our paper is structured as follows. We first point out a fundamental shortcoming of the simple estimates for the abundance of interstellar objects (\S 2). Assuming that the detected object was drawn from a reasonable size distribution implies an orders of magnitude increase in this abundance.  We then (\S 3) use existing simulations of extrasolar planetary dynamics \citep{raymond11} to develop a framework for the possible sources of ejected planetesimals. We show that this naturally matches the nominal (lower) estimate for the abundance of interstellar objects.  We argue on dynamical grounds that it is much easier to explain 'Oumuamua as a dormant cometary nucleus rather than an asteroid. In \S 4 we show that a fraction $f \sim 1\%$ of planetesimals in our simulations underwent extremely close encounters with one or more giant planets before being ejected. These encounters should lead to tidal disruption. We demonstrate that this solves the problem pointed out in \S 2 if the population of ejected planetesimals is characterized by a standard size distribution with a small contamination of fragments, which contain a small fraction of the mass but dominate by number. In \S 5 we compare our model with an alternate one that assumes 'Oumuamua to have originated in a binary star system, and conclude with a discussion of the implications for the birth size distribution of planetesimals in \S 6.

\section{Estimating the planetesimal mass ejected per star}

The observation of a flux of interstellar objects passing through the inner Solar System directly constrains the interstellar number density $n$ of such bodies. With an estimate of the size and density of the intruders, we can convert the number density to a mass density and infer the average amount of mass that must be ejected for each Solar mass of local stellar population. This estimate has been made by several authors~\citep{trilling17,laughlin17,meech17,portegieszwart17,rafikov18,do18}. It is necessarily uncertain for so long as we have only a single detected object.

Recapping the elementary argument, the cross-section for an object with relative velocity at infinity $v_\infty$ to approach to a peri-center distance $q$ is,
\begin{equation}
 \sigma = \pi q^2 \left[ 1 + \left( \frac{v_{\rm esc}}{v_{\infty}} \right)^2 \right],
\end{equation}
where $v_{\rm esc}$ is the escape velocity from $q$. The observation of a rate of such encounters $\Gamma$ 
determines the number density $n = \Gamma / \sigma v_\infty$ and an interstellar mass density,
\begin{equation}
 \rho_{\rm ej} = \frac{4 \pi \Gamma r^3 \rho_m}{3 \sigma v_\infty},
\end{equation}
where $r$ is the radius and $\rho_m$ the material density of the objects. 'Oumuamua has $v_\infty = 26 \ 
{\rm km \ s}^{-1}$ and $q = 0.25 \ {\rm AU}$, and hence we adopt $\sigma \sim 3 \ {\rm AU}^2$. Large 
variations in the absolute magnitude suggest a highly elongated ellipsoidal figure, with an equivalent 
spherical radius of $r \approx 70 \ {\rm m}$ \citep{meech17,jewitt17,bannister17,bolin18}. The lack of evidence for outgassing 
means that the purely observational prior favors an asteroid-like composition, with $\rho_m \approx 3 \ {\rm g \ cm}^{-3}$. 
The rate is highly uncertain in the absence of a detailed study of survey completeness, but is of the order of 
$\Gamma \sim 0.1 \ {\rm yr}^{-1}$.

We can compare $\rho_{\rm ej}$ to the local mass density of stars. \citet{mckee15} infer a local surface density 
of stars and stellar remnants $\Sigma_* = 33.4 \pm 3 \ M_\odot \ {\rm pc}^{-2}$, and a volume density $\rho_* = 0.060 \ M_\odot \ {\rm pc}^{-3}$. If we make the simple (and probably incorrect) assumption that the amount of mass ejected scales linearly with $M_*$, we find that the mass ejected per Solar mass of the local stellar population is,
\begin{eqnarray}
 \epsilon_{\rm ej} \simeq 0.65 \left( \frac{\Gamma}{0.1 \ {\rm yr}^{-1}} \right) 
 \left( \frac{\sigma}{3 \ {\rm AU}^2} \right)^{-1} 
 \left( \frac{v_\infty}{26 \ {\rm km \ s}^{-1}} \right)^{-1} \nonumber \\
 \times \left( \frac{r}{70 \ {\rm m}} \right)^3 
 \left( \frac{\rho_m}{3 \ {\rm g \ cm}^{-3}} \right) M_\oplus M_\odot^{-1}.
\label{eq_mass_ej} 
\end{eqnarray}
This estimate drops to $ \epsilon_{\rm ej} \approx 0.2 \ M_\oplus M_\odot^{-1}$ for an icy (lower-density) composition, and has a large uncertainty inherent in deriving a rate from a single object in an incompletely characterized survey. We take $1 \ M_\oplus M_\odot^{-1}$ as a central estimate, and consider 0.1-10~$M_\oplus M_\odot^{-1}$ as a reasonable range. This is somewhat lower than the values inferred by \cite{trilling17} and \cite{do18}, who found roughly $20 \mearth$ and $40 \mearth$ ejected per star, respectively. As we show below and in \S 4, any inferred values (including ours) are unlikely to represent the true value but rather may be underestimated or overestimated by a significant amount, subject to unconstrained parameters of the population of interstellar objects.

%\section{Constraints on the mass function of planetesimals}
%As we will see in Section 4, the ejection of $\sim 1 \ M_\oplus$ of planetesimals per Solar mass of stellar population is dynamically feasible. 

The mass estimate in equation~(\ref{eq_mass_ej}) is valid if the population of interstellar planetesimals is mono-disperse. If the mass distribution of the ejected population is instead top heavy---as is observed for the asteroid and Kuiper Belts, and expected theoretically in some planetesimal formation scenarios~\citep[see][]{johansen14}---then the inferred $\epsilon_{\rm ej}$ increases.

We model the mass function of interstellar planetesimals as a power-law,
\begin{equation}
 \frac{{\rm d}N}{{\rm d}m} = k m^{-p},
\end{equation}
between $m_{\rm min}$ and $m_{\rm max} \gg m_{\rm min}$. We assume that the mass function is dominated by number by small objects, and by mass by large ones ($1 < p < 2$). The expected mass for the first object to be seen from this distribution is given by setting the {\em number} of objects with a lower mass equal to the number with a higher mass, $m_{1/2} = 2^{1/(p-1)} m_{\rm min}$. Proceeding as above we would then estimate a total mass for the population of $M_{\rm est} = N\,m_{1/2}$, where $N$ is the total number of bodies. The true mass, however, is given by integrating over the mass function as $M_{\rm total} = k m_{\rm max}^{2-p} / (2-p)$. The ratio between the true mass of the population and the one estimated using a single size is then,
\begin{equation}
 b = 2^{p-1} \left( \frac{p-1}{2-p} \right) \left( \frac{r_{\rm max}}{r_{\rm min}} \right)^{3 (2-p)},
\end{equation}
written in terms of the minimum and maximum radii.

This leads to a discrepancy with models in which planetesimals form via the streaming instability \citep{youdin05}. Simulations show that a single burst of planetesimal formation from initial conditions unstable to streaming yields a power-law mass function with $p \simeq 1.6$ \citep{johansen15,simon16,simon17,schafer17}. Taking $r_{\rm min} = 100 \ {\rm m}$ and $r_{\rm max} = 100 \ {\rm km}$ (a conservative upper size limit, as forming a belt with a total mass of $\sim 10 \ M_\oplus$ in one episode would probably yield a maximum size closer to 1000~km), we find $b \sim 10^4$.  A still substantial bias $b > 10^2$ occurs for steeper but still top-heavy mass functions, for example for $p=11/6$ as appropriate for a steady-state collisional cascade~\citep{dohnanyi69}.

If ejected planetesimals sample a streaming mass function ($p = 1.6$), then the estimate of $1 \mearth \msun^{-1}$ in ejected planetesimals is underestimated by a factor of $b \ge 100$. The inferred mass is hundreds of Earth masses, larger than the total solid budget of most planet-forming disks (see \S\ref{subsec_limits}).  

We conclude this section with a conundrum. An estimate for the mass in interstellar planetesimals based on an unrealistic, mono-disperse size distribution yields a reasonable value.  Yet an estimate based on a plausible size distribution produces an unrealistically-large value.

\section{Dynamical sources of ejected planetesimals}
%\subsection{A baseline model}

The mass of ejected planetesimals is determined by the initial configuration of a planetary system and by the location and masses of planetesimal belts. For gas giants we have reasonable estimates of planetary frequency out to 5-10~AU~\citep{cumming08,mayor11,wittenmyer16}, and circumstantial evidence from orbital eccentricities for their dynamical history~\citep[e.g.][]{adams03,chatterjee08}. There is no evidence from direct imaging surveys or microlensing for a substantial population of gas giants in wide orbits~\citep{meshkat17,mroz17}. We know from the existence of debris discs \citep{wyatt08} that extrasolar planetesimal belts exist, but have only limited knowledge of their initial masses. 

The above constraints do not allow us to uniquely predict a stellar population averaged mass ejection efficiency. What we can do, however, is to define a set of dynamical models that fit the basic known properties of the giant extrasolar planet population, using assumptions for the masses of planetesimal belts that are in the range inferred for the Solar System. We first ask whether such a baseline model, developed previously \citep{raymond10}, is consistent with the observation of 'Oumuamua. We assume that,
\begin{itemize}
%\item
%All ejection occurs from single-star systems. In binary star systems, ejection may be more efficient \citep{smullen2016,sutherland2016,gong2017}
%\item
%All ejection occurs during the main sequence. Extrasolar analogues of the Oort Cloud -- formed partly during the birth of the planetary system and partly from continuous interchange with the interstellar medium -- will be ejected almost entirely during and after the star's asymptotic giant branch phase \citep{veras14b,stone15}. 
\item
A fraction $f_{\rm giant}$ of stars host giant planets. Within this subset, a fraction $f_{\rm unstable}$ 
form planets in an initial configuration that is subject to subsequent dynamical instability (leading to 
orbit crossing, planetary ejections, and efficient ejection of neighbouring planetesimals).
\item
The remaining stars do not host giant planets. These systems might host ice giants, but without 
secure knowledge of the ice giant populations it is reasonable to assume that they are less efficient at ejecting planetesimals on average.  
\end{itemize}

The mean mass in planetesimals ejected per star is then,
\begin{eqnarray}
M_\mathrm{interstellar} = (1 - f_\mathrm{giant}) \times M_\mathrm{ejected}^\mathrm{no giant} + \nonumber \\
f_\mathrm{giant} \times (1 - f_\mathrm{unstable}) \times M_\mathrm{ejected}^\mathrm{giant,stable} + \nonumber \\
 f_\mathrm{giant} \times f_\mathrm{unstable} \times M_\mathrm{ejected}^\mathrm{giant,unstable} 
\end{eqnarray}
\noindent where $M_\mathrm{ejected}$ is the average mass ejected from systems without a gas giant (nogiant), and in stable (giant,stable) and unstable (giant,unstable) giant planet systems.  

%Some of the parameters in the above expression can be estimated. 
Long-duration radial velocity surveys constrain the occurrence rate of giant planets $f_\mathrm{giant}$ to be $\sim$10-20\% for Sun-like stars~\citep{cumming08,mayor11,wittenmyer16,rowan16,foremanmackey16}. Most gas giants are located beyond 0.5-1 AU at radii where they could plausibly eject planetesimal belts. Low-mass stars have fewer gas giants than high-mass stars~\citep{johnson07,lovis07,dressing15} so the population-averaged rate is likely smaller than the value for Sun-like stars. 

%\begin{figure}
%%  \leavevmode \epsfxsize=8cm\epsfbox{Mejunst-Mejst.eps}
%  \leavevmode \epsfxsize=8cm\epsfbox{Mejunst-Mejst_lowmass.eps}
%    \caption[]{Mass ejected per stable/unstable giant planet system to reach $M_\mathrm{interstellar} = 0.1 - 100 \mearth$.  The hatched areas show the allowed range for our assumed parameters ($f_\mathrm{giant} = 0.01-0.1$, $f_\mathrm{unstable} = 0.75-0.95$) and the solid curves are fiducial values ($f_\mathrm{giant} = 0.03$, $f_\mathrm{unstable} = 0.9$). We have assumed that systems without giant planets do not eject planetesimals. The region below the curves is allowed if systems without giant planets do eject planetesimals.  } 
%     \label{fig:params}
%\end{figure}

Unstable giant planet systems are extremely efficient at ejecting planetesimals. Planet-planet scattering is the leading model to explain the eccentric orbits of giant exoplanets~\citep[e.g.][]{lin97,adams03,chatterjee08} and the observed distribution can be matched if $\sim$75-95\% of giant exoplanets are the survivors of dynamical instabilities~\citep[][]{juric08,raymond11}. Simulations show that instabilities typically result in the ejection of one or more giant planets and the gravitational perturbations during this process often ejects the entire outer planetesimal disc out to $\sim 30$~AU~\citep[]{raymond10,raymond11,raymond12,raymond13b,marzari14}.  

A baseline model for the source population of interstellar objects invokes efficient ejection from a small fraction of stars with unstable giant planet systems, less efficient ejection for stars with stable giants, and no ejection from the bulk of stars without massive planets (based on the Safronov number argument). Given that we infer a high unstable fraction ($f_{unstable} \approx 90\%$) from dynamical arguments, the key unknown is the mass ejected per unstable system ($M_\mathrm{ejected}^\mathrm{giant,unstable}$). The ejection of 10-100~$M_\oplus$ per star hosting unstable gas giants suffices to match the central estimate (of $1 \mearth$ ejected per Solar mass of stars) for the population averaged ejection efficiency assuming a stellar mass-averaged $f_\mathrm{giant}$ of 1-10\%. This is consistent with Solar System expectations if 'Oumuamua formed as an icy body (see \S 3.1).

It is worth noting that equation (6) likely underestimates planetesimal ejection. For example, in binary star systems, ejection may be more efficient than in single-star systems \citep{smullen2016,sutherland2016,gong17}.  Planetesimal ejection may also be efficient after the main sequence~\citep[see][for a review of the effects of stellar evolution on orbital dynamics]{veras16}. Extrasolar Oort Cloud analogues -- formed partly during planet formation and partly from continuous interchange with the interstellar medium -- are almost entirely ejected during and after a star's asymptotic giant branch phase \citep{veras14b,stone15}.\footnote{Two new studies explore a post-main sequence ejection origin for 'Oumuamua~\citep{hansen17,rafikov18}.}

\subsection{An interstellar asteroid or comet?}
Observations of 'Oumuamua have not conclusively determined its composition. Despite its very close approach to the Sun (perihelion of 0.25 AU), 'Oumuamua has shown no signs of sublimed water in the form of a coma~\citep{meech17,ye17,jewitt17}. However, \cite{fitzsimmons18} showed that 'Oumuamua's perihelion passage would only have exposed volatiles down to a depth of roughly 40 cm.  'Oumuamua's colors fall within the distribution of primitive, organic-rich Solar System objects~\citep{jewitt17,bannister17,fitzsimmons18,bolin18}. It has a red spectral slope with no obvious absorption features~\citep{meech17,masiero17,ye17,fitzsimmons18}, similar to some primitive Solar System bodies~\citep[e.g.][]{bus02,carry16}.  Put together, it seems plausible to imagine that 'Oumuamua is either rocky (asteroidal) in nature or an extinct cometary nucleus (as we discuss further below). The Damocloids, with orbits similar to Halley-type and long-period comets but no detectable activity~\citep{asher94,jewitt05}, may offer a suitable analogy. On the other hand, \cite{cook16} found that surveys may be biased toward finding asteroids if small ($r < 1$~km) comets are very rare.

\begin{figure}
%  \begin{center} 
 \leavevmode \epsfxsize=8cm\epsfbox{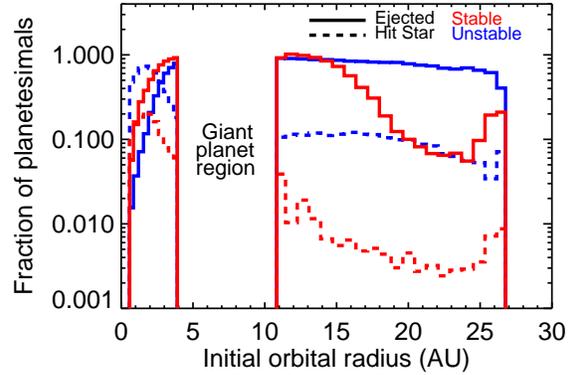}
    \caption[]{The fate of planetesimals in the simulations of \cite{raymond11}.  The solid [dashed] lines correspond to the fraction of planetesimals with a given starting orbit that were ejected [hit the central star]  in systems in which the giant planets' orbits remained stable (red) or went unstable (blue). These simulations assumed solar-mass stars so all planetesimals from exterior to the ice giants' orbits are likely to be cometary.} 
\label{fig:ejec}
\end{figure}

Given the inconclusive direct evidence we use a suite of dynamical simulations to address the question of whether asteroidal or cometary planetesimals are more likely to be ejected from systems with gas giants. The Safronov number for a planet of given mass increases with orbital radius, so from a dynamical standpoint it is easier to eject icy planetesimals from beyond the snow line than rocky bodies that form closer to the star~\citep[e.g.,][]{gaidos17}.\footnote{A few per cent of rocky bodies may be scattered into outer icy regions where they did not form \citep{weissman1997,shannon2015,meech16}, but this does not affect our argument.}

Figure~\ref{fig:ejec} shows the fraction of planetesimals that were ejected or hit the central star in simulations from \cite{raymond11,raymond12}. These simulations included an inner disc of terrestrial-planet forming (rocky asteroidal) material, three giant planets (with masses chosen to match the observed planetary mass function), and an outer disc of presumably comet-like, volatile-rich planetesimals. In these simulations, dynamical instabilities in the (dominant) subset of initially unstable systems were almost 100\% efficient in ejecting icy planetesimals across a broad range of radii. For the rocky planetesimals, on the other hand, a substantial fraction were excited to high enough eccentricities that they impacted the star. The same trends are seen in the stable subset~\citep[see also][]{barclay17}, but these systems were only able to eject nearby planetesimals in the outer belt.

The above arguments do not rule out the possibility that the majority of interstellar bodies are asteroidal in nature. This would require, however, that primordial rocky planetesimal belts are typically substantially more massive than icy ones at larger radii. Moreover, the frequency of massive planets that could eject asteroids from close-in orbits is already constrained. Roughly, we would require typical asteroid belts of the order of $10 \ M_\oplus$ to meet the central estimate for ejection efficiency. This is not impossible, since aerodynamically coupled pebbles may drift inward and preferentially produce planetesimals close-in~\citep{drazkowska16}. We would be forced to consider the possibility that asteroid belts are typically more massive than belts of icy planetesimals if future observations suggest that most interstellar bodies are rocky.

%The simulations shown in Figure~\ref{fig:ejec} modeled ejection of planetesimals from an isolated star, and did not include any tidal effects at large radii. The ejections were therefore prompt. In a more realistic scenario, some fraction of the planetesimals would have ended up in an extrasolar analogue of the Oort cloud. This reservoir, however, would be liberated from the fraction of stars that exit the main sequence and lose mass during the asymptotic giant branch phase within the lifetime of the Galaxy \citep{veras14,stone15}.

\subsection{Limits on ejected masses}
\label{subsec_limits}
A planetary system cannot eject an arbitrarily large amount of planetesimals because the process is self-limiting.  A growing gas giant ejects a significant fraction of nearby planetesimals~\citep{raymond17} and, after dispersal of the gaseous disc, planetesimals from within a few Hill radii are perturbed onto giant planet-crossing orbits and ejected~\citep{duncan87,charnoz03}.  At this point ejection is slowed because the reservoir of unstable planetesimals is cut off~\citep[e.g.][]{bonsor12}. Additional planetesimals can be ejected if a giant planet is perturbed onto an eccentric orbit, usually by planet-planet scattering. However, if the mass in outer disc of planetesimals is within a factor of a few of the scattered planet's mass, angular momentum exchange damps the planet's eccentricity and lifts its pericenter, quenching the giant planet instability~\citep{raymond09b,raymond10}. For this reason we do not expect a system to eject more than its planetary budget in planetesimals~\citep[at least while its host star is on the main sequence; see][]{veras14b,stone15,rafikov18}.

Planetesimal formation could in principle start early, when the disc mass is high. For a gas disc mass of $0.1 \ M_\odot$, the mass in icy material for a standard metallicity is approximately 300~$M_\oplus$. If extremely massive planetesimal belts formed frequently, the easiest way to eject them without violating observational constraints on the number of planets at large orbital radii would be if the same systems hosted multiple ice giants. In such a model of the order of $100 \ M_\oplus$ of icy planetesimals could be ejected from most stars. This scenario appears contrived, though it shares some similarities with models which seek to explain the dust rings in systems such as HL~Tau \citep{alma15} as a consequence of early-forming ice giants.
 
\section{Consequences of tidal disruption during ejection}

During the scattering phase, planetesimals undergo close encounters during which they pass within a gas giant's Hill sphere. Close encounters on the pathway to ejection can tear planetesimals apart. It is worth considering whether 'Oumuamua could itself be a fragment of a larger planetesimal that was tidally disrupted. If disrupted ejected planetesimals are to be common, they must walk a tightrope: too close an approach to a giant planet will result in a collision, and many distant approaches will not cause disruption.

We investigated the encounters undergone by planetesimals on the path to ejection. We used the same batch of simulations as presented in Fig.~\ref{fig:ejec}~\citep[from][]{raymond11,raymond12}. For each of the $>10^5$ planetesimals ejected in those 200 simulations, we tabulated the parameters of the close encounters undergone with each of the three giant planets before the planetesimal's ejection. We divided up the planetesimals by origin; planetesimals closer to the star than the giant planets (between 0.5 and 4 AU) are referred to as `asteroidal' and planetesimals originating beyond the giant planets (typically past 10 AU; see Fig.~\ref{fig:ejec}) are `cometary'.  In contrast with Fig.~\ref{fig:ejec}, we now combine both stable and unstable simulations into a single set.

While comets were ejected at a higher rate than asteroids (Fig.~\ref{fig:ejec}), ejected asteroids typically underwent more encounters than ejected comets. The median number of close encounters with a giant planet before ejection was 54 for asteroids and 19 for comets. There is a clear correlation between the number of encounters and the closest close encounter: planetesimals that underwent more encounters had a statistically closer closest encounter.

The critical radius for tidal disruption $R_t$ is defined as\footnote{We have deliberately kept our formulation for $R_t$ simple; more complex treatments would take into account the effect of other physical properties of planetesimals including tensile strength~\citep[e.g.,][]{cordes08,veras14,bear15,veras16}.}
\begin{equation}
R_t \approx r \left(\frac{M_{giant}}{m}\right)^{1/3},
\end{equation}
\noindent where $m$ and $r$ are the planetesimal's mass and radius, respectively, and $M_{giant}$ is the giant planet mass. It is clear that $R_t \propto \rho^{-1}$, where $\rho$ is the planetesimal's bulk density.  For a planetesimal with density of 1~g~cm$^{-3}$ undergoing a close encounter with Jupiter, $R_t \sim 76,800$~km, just $\sim 10\%$ wide of Jupiter's actual radius of $\sim 70,000$~km. 

\begin{figure}
  \leavevmode \epsfxsize=8cm\epsfbox{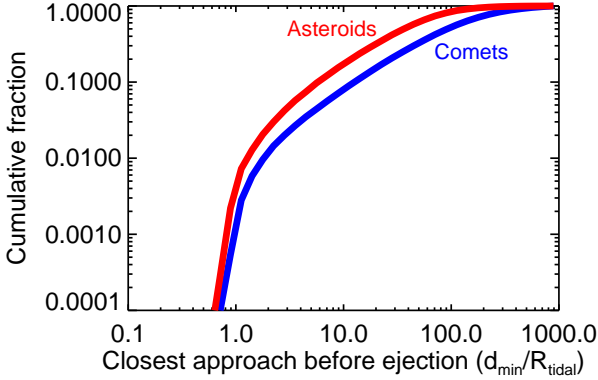}
%  \leavevmode \epsfxsize=8cm\epsfbox{dminRT_astcom_notext.eps}
    \caption[]{Cumulative distribution of the closest close encounter undergone by planetesimals in the process of being ejected. The closest encounter distance $d_{min}$ is normalized to the tidal disruption radius $R_t$ for that planet, assuming for the calculation of $R_t$ that all planetesimals have bulk densities of 1 g cm$^{-3}$.  The red 'asteroid' curve corresponds to planetesimals that started the simulations interior to the giant planets' orbits, and the blue 'comets' curve are planetesimals originating beyond the giant planets' orbits~\citep[in the simulations from][]{raymond11}. } 
     \label{fig:encounters}
\end{figure}

Figure~\ref{fig:encounters} shows the cumulative distribution of the closest encounter for ejected asteroids (in red) and comets (in blue), scaled to the tidal disruption radius $R_t$ of the relevant planet.  The median closest encounter was $35 R_t$  for asteroids and $87 R_t$ for comets.  The fraction of planetesimals that underwent encounters within 1, 3 and 10 $R_t$ was, respectively, 0.2\%, 3.9\%, and 16\% for asteroids and 0.05\%, 1.9\%, and 7\% for comets. Among planetesimals with approaches closer than $R_t$, 18\%/6\%/1.4\% had 2/3/4 or more passages within $R_t$, i.e., multiple likely-disrupting encounters.

The kink in the distribution of closest encounters at $d_{min} \le R_t$ is due to the tidal radius approaching the planet's physical radius such that many encounters within $R_t$ simply lead to planetary collisions, not ejections.  Given that giant planets compress and shrink in radius as a function of time~\citep[e.g.][]{baraffe03,fortney10}, the parameter space available for tidal disruption followed by ejection (rather than a simple collision) increases at later times. In our simulations each of the giant planets -- with masses drawn from the observed $dN/dM \propto M^{-1.1}$ distribution~\citep{butler06,udry07b} between Saturn's mass and three Jupiter masses -- was given Jupiter's bulk density of 1.3 g cm$^{-3}$. This may underestimate the radius of some planets and modestly overestimate the contribution of tidally disrupted ejected planetesimals.

What happens when a planetesimal passes inside the tidal disruption radius?  One would expect the object to be torn apart down to the size scale at which tensile strength dominates over self-gravity, orders of magnitude smaller than the typically-assumed planetesimal size of $\sim 100$~km~\citep[e.g.][]{benz99,obrien03,leinhardt09}. This would create a huge number of tidally-distorted fragments, possibly with large axis ratios like 'Oumuamua's~\citep{meech17,jewitt17,bolin18,bannister17,fraser17,fitzsimmons18}. After tidal disruption, the resulting debris could recollapse into larger bound objects \citep{hahn98,veras14,coughlin16}.  Detailed modeling of this process is beyond the scope of this paper, yet it seems reasonable to assume that a very large number of small fragments may be created by the disruption of a single large planetesimal. Indeed, comet Shoemaker-Levy 9's close encounter with Jupiter in 1992 caused it to disrupt into a chain of 21 detectable fragments~\citep{weaver95,noll96,movshovitz12}, many of which were similar in size to 'Oumuamua.  

We now explore the consequences of a bimodal population of interstellar planetesimals. We naively assume that a small fraction $f$ from a dN/dM=k$\,$m$^{-p}$ profile has been tidally disrupted and transformed into mono-sized fragments with size $r_{frag}$ and mass $m_{frag}$ (assuming a constant physical density across all sizes). The total number of fragments $N_{frag}$ is
\begin{equation}
N_{frag} = f \left(\frac{k}{2-p}\right) \left(\frac{m_{max}^{2-p}}{m_{frag}}\right),
\end{equation}
assuming that $1 < p < 2$ such that the mass is dominated by the largest bodies. Given that the constant $k$ can be written in terms of the total number of bodes in the initial distribution $N$ (as $k \approx \frac{N}{p-1}m_{min}^{1-p}$) the ratio of the total number of fragments to the bodies in the original distribution is,
\begin{equation}
\frac{N_{frag}}{N} \approx f \left(\frac{p-1}{2-p}\right) \left(\frac{m_{max}^{2-p}}{m_{min}^{1-p}}\right)\left(\frac{1}{m_{frag}}\right).
\end{equation}
For fiducial values let us assume an initial planetesimal population that follows a streaming instability distribution with $p=1.6$ from $r_{min} =$~1~km up to $r_{max}=$~100~km with bulk densities $\rho = 1\ {\rm g \ cm}^{-3}$. We further assume that, as in our simulations (Fig.~\ref{fig:encounters}), $f = 1\%$ of planetesimals are disrupted and produce fragments of $r_{frag} = 100$~m in size (similar to 'Oumuamua).  With those assumed values, the number of fragments dominates the total distribution, with $N_{frag}/N$ of a few thousand.  If we instead assume that the fragments created match the minimum in the already-established size distribution, i.e., that $r_{frag} = r_{min}$, then $N_{frag}/N \approx 4$.  

In Section 2 we showed that assuming a mono-size distribution for all interstellar objects yielded an underestimate of the true mass. This was quantified in Eq. 5 using the bias $b$, which was typically two or more orders of magnitude. This discrepancy is alleviated in a natural fashion if $N_{frag}/N \sim b$, that is, fragments make up the dominant component of interstellar objects. The balance between the contribution from fragments and the bias can be written simply as
\begin{equation}
\frac{N_{frag}/N}{b} = \frac{f}{2^{p-1}} \left(\frac{r_{min}}{r_{frag}}\right)^3.
\end{equation}
If $N_{frag}/N>b$ then fragments dominate the population of interstellar objects to a sufficient extent that the mono-disperse value from equation 3 overestimates the true value. If $N_{frag}/N<b$ the mono-disperse value still underestimates the true value.

Both an underestimate and an overestimate are entirely within reason. For plausible choices of parameters -- $p=1.6$, $f=1\%$, $r_{min}$~= 1 km, and $r_{frag}$ = 100 m -- the mass inferred from a mono-sized distribution overestimates the true one by a factor of 6.6. This would reduce \cite{trilling17}'s and \cite{do18}'s estimate to $3 \mearth$ and $6 \mearth$ in planetesimals ejected per star, respectively.  For only slightly different parameters -- keeping $p=1.6$ and $r_{min}$~= 1 km but decreasing $f$ to 0.5\% and increasing $r_{frag}$ to 200 m -- the mass inferred from a mono-sized distribution {\em underestimates} the real one by a factor of 2.4. This would increase \cite{do18}'s estimate to $\sim 100 \mearth$ ejected per star. Given the large uncertainty in parameters it is premature to over-interpret these inferred masses.

It is interesting that our nominal parameters come close to entirely canceling out the bias. If we naively equate $N_{frag}/N$ and $b$ we can solve for the fraction of planetesimals that must disrupt on the way to ejection to perfectly cancel each other out:
\begin{equation}
f = 2^{p-1}\left(\frac{r_{frag}}{r_{min}}\right)^3.
\end{equation}
Assuming that fragments are much smaller than the smallest planetesimals that form by the streaming instability, i.e., $r_{frag} = 0.1 \ r_{min}$, only 0.15\% of ejected planetesimals need to perfectly cancel out the bias. If fragments are slightly larger -- $r_{frag} = 0.2 \ r_{min}$ -- then $f \approx$~1\%, as we found in our simulations (Fig.~\ref{fig:encounters}).  

%This concordance is powerful. 
If roughly 1\% of planetesimals undergo tidal disruption events on their way to ejection, and if the typical fragment size is modestly smaller than the smallest bodies in the initial distribution, then the large difference (quantified by $b$ in Eq. 5) between estimates of the mass in interstellar planetesimals for mono-disperse and realistic size distributions decreases significantly and may even disappear. This would imply that the distribution of interstellar objects is dominated by fragments by number and by the largest planetesimals by mass. If this interpretation is correct we would expect future interstellar objects to be dominated by objects similar in size to 'Oumuamua and also to show signs of tidal distortion via their stretched-out shapes. Alternately, if this interpretation is wrong and the true distribution is similar to the streaming one, then we would expect a much lower rate of future discoveries than predicted~\citep[e.g., by][]{trilling17}.

Finally, at face value Fig.~\ref{fig:encounters} shows that asteroidal planetesimals are disrupted before ejection at a higher rate than cometary planetesimals. Given the potential for a large number of fragments per disruption event, a relatively small difference in efficiency could create a large discrepancy by number. This might tilt the scales in favor of detecting asteroidal interstellar planetesimals rather than cometary ones, in contrast with the arguments presented in \S 3.1.  However, this interpretation remains premature. Asteroids have higher physical densities than comets~\citep{carry12}, with correspondingly smaller tidal disruption radii. The shrinking value of $R_t$ would shift the asteroid curve to the right in Fig.~\ref{fig:encounters}, toward the cometary curve. For factor of two higher physical density for asteroids, the difference in disruption efficiency between ejected asteroids and comets vanishes. In fact, it may be that disruption is extremely rare for asteroids.

\section{Comparison with other origin hypotheses for 'Oumuamua}

We have argued that single stars with gas giants can account for the abundance and properties of 'Oumuamua. In our scenario, 'Oumuamua is likely to be the fragment of an extinct comet that was born beyond the snow line of its parent star.\footnote{Several studies have tried to pinpoint the Galactic origins of 'Oumuamua but there is currently insufficient evidence to clearly link its trajectory to a specific star or star-forming region~\citep{mamajek17,gaidos17,zuluaga17,zhang18}.}
Planetesimals will also be ejected (and may be tidally disrupted) from binary star systems. Binary systems are abundant in the Galaxy~\citep{duchene13}, and a binary companion  
provides stronger dynamical perturbations than a gas giant. Typically, however, we would expect a binary to interact with a much smaller fraction of surrounding planetesimals than an (initially unstable) giant planet system, and hence there is a trade-off which must be assessed quantitatively in order to determine if binary ejections are significant. For completeness, we briefly discuss two other origin hypotheses for 'Oumuamua that invoke binary systems.

\cite{jackson17} proposed that most interstellar planetesimals form in circumbinary disks. Close binary stars have well-defined dynamical stability limits, as objects that enter within a critical orbital radius are destabilized~\citep{holman99}. \cite{jackson17} assume that planetesimals form beyond the stability limit and systematically drift inward across the limit under the effect of aerodynamic gas drag. They performed simulations to show that all planetesimals that drift inside the stability limit are ejected. For a significant fraction of close binaries the snow line is exterior to the stability limit such that some ejected planetesimals may be volatile-poor. This process could dominate the population of interstellar planetesimals if two pre-conditions are met. First, a sufficiently large fraction of the entire solid disc mass must drift interior to the binary stability limit (\cite{jackson17} assume this fraction is 10\%). If the drift occurs while the solids are in the form of planetesimals, this requires relatively small planetesimals ($r \le$~1 km) which can drift significantly under aerodynamic forces during the disc lifetime \citep[e.g.,][]{adachi76,thommes03}. Second, the drifting planetesimals need to enter the dynamically unstable zone close to the binary, rather than piling up at the pressure maximum in the gaseous circumbinary disc which may be at modestly greater orbital radii \citep{artymowicz94,pierens07,marzari08,lines16}.

\cite{cuk18} proposed that 'Oumuamua is a fragment of a planet that formed in a binary system and was tidally disrupted after passing too close to one of the stars, which itself must be sufficiently dense to allow for tidal disruption rather than collision (this favors M dwarf stars). This would represent a low-probability event but one that creates enough fragments, if ejected, to flood the distribution of interstellar planetesimals. \cite{cuk18} suggests that these events are possible in S-type binary systems, in which the binary companion is on a wider orbit than the planet. It is likely, however, that planet-planet scattering provides one of the dominant channels for triggering planet-star near-collisions, and hence the `direct' planetesimal ejection / tidal disruption mechanism that we have discussed would necessarily accompany planetary disruptions. It is also unclear whether planetary tidal disruptions would yield fragments of the size inferred for 'Oumuamua.

\section{Conclusions}

The observation of 'Oumuamua provides a first crude estimate of the number density of interstellar planetesimals. Converting that number density to a mass is model-dependent and highly uncertain. 
By assuming a mono-disperse source population, we calculated a seemingly reasonable value of $1 \ M_\oplus$ in ejected planetesimals per stellar mass in stars (\S 2).  However, for a mass function that is dominated by number by small bodies, but by mass by large ones, the first observation will almost always be of an object close to the minimum size. Inferring an interstellar mass density from the derived number density then leads to an underestimate of the typical ejected mass, typically by several orders of magnitude. Future observations of interstellar objects therefore have the potential to place unique constraints on the mass distribution of planetesimals, with the caveat that the ejected bodies do not necessarily fairly sample the primordial distribution.

Dynamical considerations suggest that systems with gas giants are responsible for ejecting the bulk of interstellar planetesimals (\S 3).  Cometary planetesimals from beyond the gas giants' orbits are ejected with a much higher efficiency than asteroidal planetesimals from closer-in.  Ejection is most efficient in systems in which the gas giants themselves become unstable and undergo a phase of planet-planet scattering, thus ejecting the bulk of outer planetesimal disks~\citep[see Fig.~1 and ][]{raymond11,raymond12}.  

During the process of ejection, some planetesimals may pass so close to giant planets that they are tidally disrupted (\S 4). This process requires planetary densities that are comparable to or higher than (present-day) Jupiter, and hence is sensitive to the abundance of super-Jovian planets and to the timing of planet-planet scattering. However, it appears possible that of the order of 1\% of planetesimals could be tidally disrupted during ejection. If those planetesimals are shredded into much smaller ('Oumuamua-sized) fragments, then the fragments dominate by number (but not by mass). For reasonable parameter choices, the mass inferred for a mono-disperse interstellar population approaches the true mass of a more complex distribution, although the mono-disperse mass may still under- or over-estimate the true mass by a large factor (see Eq. 10). If 'Oumuamua represents a fragment of a disrupted planetesimal, then the disruption event itself may be the key to explaining its shape and tumbling rotation~\citep{fraser17,drahus17}.  

%If the majority of interstellar objects turn out to be asteroidal in nature the conclusion is more interesting. Ejection of asteroids is disfavoured from both a dynamical and planetesimal formation standpoint. A dominant population of asteroids would imply large masses for extrasolar asteroid belts, and a small ratio by number of icy to rocky bodies.

%Our interpretation comes with some caveats. For instance, we cannot rule out that more realistic simulations of the streaming instability in the future may yield a steeper mass function. Alternatively, one may appeal to a delay between the formation of planetesimals and the formation (or migration) of the ejecting giant planets. It is possible to imagine a scenario in which planetesimals form with a top-heavy mass function, and then collisionally evolve to smaller sizes and drift radially (under aerodynamic forces) into dynamical contact with ejecting planets. This would break the dynamical indifference of the ejection process to mass, and allow the ejected population to have a distinct size distribution from that of the planetesimal reservoir. Finally, icy planetesimals may form from coagulation rather than streaming, or both processes may coexist and generate a bimodal population.

\section*{Acknowledgments}
We thank the referee for a helpful report. We acknowledge discussions with Arnaud Pierens, Franck Selsis, Jean-Marc Hur{\'e} and Jeremy Leconte, and helpful comments from Karen Meech, Robert Jedicke, Darin Ragozzine, and Jorge Zuluaga.  D.V. acknowledges the support of the STFC via an Ernest Rutherford Fellowship (grant ST/P003850/1). P.J.A. acknowledges support from NASA through grant NNX16AB42G.
S.N.R. thanks Agence Nationale pour la Recherche grant ANR-13-BS05-0003-002 (grant MOJO) and NASA Astrobiology Institute's Virtual Planetary Laboratory Lead Team, funded via the NASA Astrobiology Institute under solicitation NNH12ZDA002C and cooperative agreement no. NNA13AA93A.

%\bibliographystyle{mn2e_b}
%\bibliography{refs.bib}

%\begin{equation}
%M_\mathrm{interstellar} = f_\mathrm{giant} \times M_\mathrm{ejected}^\mathrm{giant} + (1 - f_\mathrm{giant}) \times M_\mathrm{ejected}^\mathrm{no giant},
%\end{equation}
%\begin{equation}
%M_\mathrm{ejected}^\mathrm{giant} = f_\mathrm{unstable} \times M_\mathrm{ejected}^\mathrm{unstable} + (1 - f_\mathrm{unstable}) \times M_\mathrm{ejected}^\mathrm{stable},
%\end{equation}

\end{document}